\documentclass{aa}  

\usepackage{graphicx}
\usepackage{txfonts}
\usepackage{lipsum}
\usepackage{upgreek}
\usepackage{subcaption}  
\usepackage{lscape}
\usepackage{placeins}

\usepackage{hyperref}
\hypersetup{
  colorlinks  = true,
  linkcolor   = blue,
  anchorcolor = blue,
  citecolor   = blue,
  filecolor   = blue,
  urlcolor    = blue
}

\defcitealias{2026NatAs..10.1020W}{W26}
\begin{document}

\title{On the triple nature of the PSR~J0435+3233 system}
    \author{
        P.~C.~C.~Freire\inst{1}\corrauth{pfreire@mpifr-bonn.mpg.de}
        \and C.~J.~Clark\inst{2,3}\corrauth{colin.clark@aei.mpg.de}
        \and C.~G.~Bassa\inst{4}\corrauth{bassa@astron.nl}
        \and G.~Voisin\inst{5}
        \and R.~van~Haasteren\inst{2,3}
        \and
        L.~Nieder\inst{2,3}
        \and B.~W.~Stappers\inst{6}
        }

   \institute{Max-Planck-Institut f\"ur Radioastronomie, Auf dem H\"ugel 69, 53121 Bonn, Germany
   \and Max Planck Institute for Gravitational Physics (Albert Einstein Institute), 30167 Hannover, Germany
   \and Leibniz Universität Hannover, 30167 Hannover, Germany
   \and ASTRON, Netherlands Institute for Radio Astronomy, Oude Hoogeveensedijk 4, Dwingeloo, 7991\,PD, The Netherlands
   \and LUX, Observatoire de Paris, Universit\'e PSL, Sorbonne Universit\'e, CNRS, 92190 Meudon, France
   \and Jodrell Bank Centre for Astrophysics, University of Manchester, M13 9PL Cheshire, UK
   }

    \date{Received XXX; accepted YYY}

    \abstract
    {The recent pulsar timing ephemeris of PSR~J0435+3233 indicates that this millisecond pulsar (MSP) has a spin-down rate that is much higher than observed in other MSPs and challenges our understanding of the formation and evolution of MSPs.}
    {We propose that this system is a hierarchical triple, and that the high spin-down rate is caused by varying acceleration due to a tertiary in a wide orbit.} 
    {We use pulsar timing methods with radio and $\gamma$-ray observations of PSR\,J0435+3233 to determine the system properties.}
    { We find that a hierarchical triple timing model describes the timing observations of PSR\,J0435+3233 and that this results in the detection of $\gamma$-ray pulsations back to the beginning of the {\it Fermi} Large Area Telescope (LAT) data in 2008. The intrinsic spin-down rate remains uncertain as it correlates with the parameters of the outer orbit, but large spin-down rates are excluded and the intrinsic rate is at least two orders of magnitude lower than the observed rate, in line with other Galactic MSPs. We identify a star located 11~mas from the pulsar position as the optical counterpart to the tertiary companion. From the 1.5-2.5\,kpc distance and colours, we infer that the tertiary is a 1.2\,M$_\odot$ F-type main-sequence star. Along with the pulsar binary, it orbits the common centre of mass with an eccentric ($e \approx 0.6$), wide ($\approx 70$~yr) orbit that is likely seen at a low orbital inclination.}
    {We conclude that PSR~J0435+3233 is a hierarchical triple system. We discuss the motivation and prospects for the continued study of this system. Spectral measurements of the outer star in addition to continued astrometric measurements will yield mass ratio and inclination estimates, while continued pulsar timing may yield a tighter constraint on violations of the Strong Equivalence Principle than are currently obtained from PSR~J0337$+$1715.}


   \maketitle

 \nolinenumbers
 
\section{Introduction}

PSR~J0435+3233 \citep[][hereafter \citetalias{2026NatAs..10.1020W}]{2026NatAs..10.1020W} is a radio pulsar with a spin period $P \simeq 3.2\, \rm ms$ discovered using the 500-m FAST radio telescope by the Commensal Radio Astronomy FAST Survey \citep[CRAFTS,][]{CRAFTS}. The spin period would generally imply that it is a millisecond pulsar (MSP), which are also known to have small spin-down rates, generally $\dot{P}< 10^{-19}$\,s\,s$^{-1}$ \citep{2005AJ....129.1993M}. The pulsar is in a $\sim 8$-d low-eccentricity orbit with a low-mass ($\gtrsim 0.3 \, \rm M_{\odot}$) companion that is likely to be a white dwarf.

What is unusual about the pulsar is the observed rate of variation of the spin period\footnote{Measured at the refence epoch, MJD = 59999},
$\dot{P} = 4.88 \times 10^{-17}$\,s\,s$^{-1}$, which is two orders of magnitude higher than those of any other known Galactic MSPs, and about three orders of magnitude larger than what is typically found among the MSP population. Taken at face value, these values of $P$ and $\dot{P}$ would imply a characteristic age of $\sim 10^6\, \rm yr$, a surface magnetic field of $1.26 \times 10^{10} \, \rm G$ and a spin-down power of $\sim 5.9 \times 10^{37} \, \rm erg \, s^{-1}$. This would imply that the pulsar is orders of magnitude younger and more powerful than any other known MSPs.  
These numbers are large enough to challenge our understanding of the formation and evolution of MSPs, and \citetalias{2026NatAs..10.1020W} attributed this to a unique formation channel for this pulsar.

Millisecond pulsars also generally emit $\gamma$-ray pulsations, with the observed population having a general trend where the $\gamma$-ray luminosity $L_{\gamma} \propto \sqrt{\dot{E}}$ \citep{2023ApJ...958..191S}. The very large $\dot{E}$ estimated by \citetalias{2026NatAs..10.1020W} should therefore imply the production of bright $\gamma$-ray emission. The {\it Fermi} Large Area Telescope \citep[LAT;][]{LAT} Fourth Source Catalogue \citep[4FGL;][]{4FGL, 4FGLDR3, 4FGLDR4} does contain a source 4FGL~J0435.5+3232 at the position of PSR~J0435$+$3233, but the energy flux from this source, $(2.4\pm0.4)\times10^{-12}$~erg~cm~$^{-2}$~s$^{-1}$ would be among the lowest for a $\gamma$-ray MSP \citep{2023ApJ...958..191S}. 

To investigate the matter further, \cite{2026arXiv260716119Z} used the timing solution of \citetalias{2026NatAs..10.1020W} to fold the $\gamma$-ray photons detected by the {\it Fermi} LAT at the position of the pulsar. They did identify $\gamma$-ray pulsations from PSR~J0435+3233, but find them to be much weaker than expected given the large $\dot{E}$ estimated by \citetalias{2026NatAs..10.1020W}. From this they conclude that the pulsar has an anomalously small $\gamma$-ray efficiency, $L_{\gamma} / \dot{E} \sim 10^{-5}$, far smaller than seen in other MSPs. Furthermore, the $\gamma$-ray pulsations were detected only in the time interval of the FAST radio observations, which starts in 2020 September 9 and ends in 2025 August 8. This is not surprising: the solution published in \citetalias{2026NatAs..10.1020W} describes the observed evolution of the spin and orbital phase via Taylor expansions, which rapidly lose predictive power outside the fitted data span.

Furthermore, motivated by the extremely high apparent spin-down power, \cite{2026arXiv260718219M} performed a targeted search for gravitational waves from PSR~J0435+3233 using the \citetalias{2026NatAs..10.1020W} pulsar ephemeris. The resulting non-detection (taking, again, the observed $\dot{P}$ at face value) rules out an explanation in which gravitational wave emission is the main mechanism through which the pulsar loses rotational energy.

In this paper, we present an alternative interpretation of the data: that the system is part of a hierarchical triple star system, one of only a handful known in the Galactic disk, including "the" Triple System PSR~J0337+1715 \citep{2014Natur.505..520R} and possibly the black-widow binary PSR~J1555$-$2908 \citep{NiederJ1555} and the eccentric MSP J1618$-$3921 \citep{2024A&A...691A..22G}. In Section~\ref{sec:derivatives}, we explore in detail the spin and orbital frequency evolution of the system. In Section~\ref{sec:timing}, we derive, from the observed frequency derivatives, a plausible orbit for the outer system, an associated phase-connected timing solution, and a refinement of the outer orbital parameters via the detection of $\gamma$-ray pulsations throughout the {\it Fermi}-LAT data. In Section~\ref{sec:third_star} we identify an optical counterpart and determine the parameters of the third body. Our results are discussed in detail in Section~\ref{sec:discussion}. We conclude with a discussion of future prospects in Section~\ref{sec:conclusions}.

\section{Spin and orbital frequency derivatives}
\label{sec:derivatives}

In addition to the large $\dot{P}$, \citetalias{2026NatAs..10.1020W} presented in their Table 1 four additional spin frequency derivatives. Both the relative and absolute magnitudes of these frequency derivatives are largely left unexamined in their paper, but the fact that five spin frequency derivatives can be measured is attributed to timing noise, which is indeed more prevalent for pulsars with large spin-down power \citep{ShannonCordesTN}.

\begin{figure}
    \centering
    {\includegraphics[width=\columnwidth]{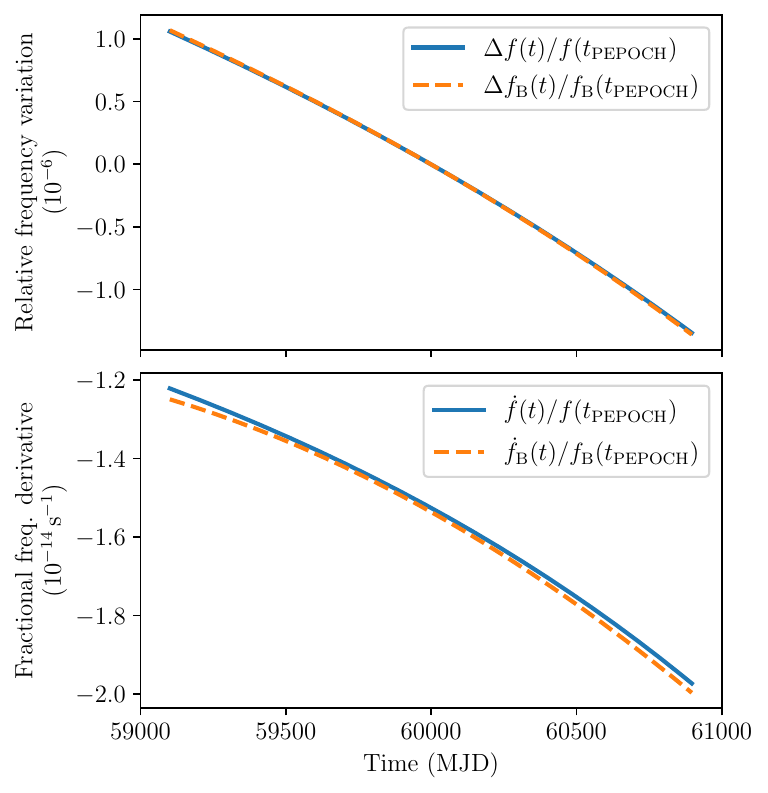}}
     \caption{Evolution of the spin and orbital frequencies (top) and spin/orbital period derivatives (bottom), divided by their values at the reference epoch $t_{\rm PEPOCH} = 59999$ (MJD), during the span of observations in \citetalias{2026NatAs..10.1020W}, according to the derivatives reported therein.}
     \label{fig:periods}
\end{figure}

In Fig.~\ref{fig:periods}, we display the evolution of the spin frequency $f$ and its first derivative $\dot{f}$ as a function of time. The curves are displayed for the observation span from \citetalias{2026NatAs..10.1020W}, using the spin frequency derivatives reported therein.

In the top plot, it is readily apparent that $f$ is not changing linearly with time, which implies a significant variation of the very large (negative) $\dot{f}$. And indeed, in the lower plot, we see how this large $\dot{f}$ is changing with time: during the 5 years of observations its magnitude increased by $\sim$ 60\%.

This raises a fundamental issue for the interpretation advanced in \citetalias{2026NatAs..10.1020W}: with changes of about 60\% in the value of $\dot{P}$ in five years, it becomes unclear how any firm conclusions can be obtained regarding the characteristic age of the system. Instead, this pronounced change of the spin period derivative suggests that the system is accelerating in the gravitational field of an additional, distant body, as demonstrated in the case of the supposed isolated pulsar PSR~J1024$-$0719 \citep{2016MNRAS.460.2207B,2016ApJ...826...86K}, which was found to be in a very wide binary, and for several binary pulsars with distant companions like PSR~B1620$-$26 in the globular cluster M4 \citep{1997ApJ...479..948J,1999ApJ...523..763T}, PSR~J0514$-$4002A in the globular cluster NGC~1851  \citep{2025A&A...697A.166D} and potentially the Galactic systems PSR~J1555$-$2908 \citep{NiederJ1555} and PSR~J1618$-$3921 \citep{2024A&A...691A..22G}.

In addition to the large spin frequency derivatives, \citetalias{2026NatAs..10.1020W} also published a total of three orbital frequency derivatives. These change the orbital frequency significantly, but the magnitude of this change is not examined in detail in their paper. 
These were attributed to orbital phase noise, common in "spider" binary pulsars (for a wide survey of their behaviour see \citealt{2025A&A...698A.239B}). If PSR~J0435+3233 were a "spider" system, its mass and orbital period would place it in the "huntsman" class (e.g. \citealt{2025ApJ...980..124S} and references therein).
Apart from the large variability in their orbital period, spider systems have radio eclipses, which are very prevalent feature in the massive spider classes (the "huntsmen" and "redbacks"). \citetalias{2026NatAs..10.1020W} report no eclipses for PSR~J0435+3233, so we find it is unlikely to be a "huntsman" binary.

In Fig.~\ref{fig:periods}, in addition to the spin frequency and its derivative, we display the change of the orbital frequency and its derivative with time during the span of observations reported in \citetalias{2026NatAs..10.1020W}. The frequencies are shown relative to their values at the reference epoch. The orbital frequency is not evolving randomly, as one would expect in a "spider" system", but it is instead evolving in a way that is almost identical to the spin frequency evolution.

This behaviour can only be adequately explained by the third distant body hypothesis, with the near-identical spin, $\dot{f}$ and orbital-frequency, $\dot{f}_{\rm B}$, evolution being due to the resultant varying Doppler shift (and nothing else). In particular, if the intrinsic spin-down and orbital variations are small (as one would expect from, for instance, a stable MSP - white dwarf system), the following relation applies 
\begin{equation}
\frac{\dot{f}}{f} \simeq  \frac{\dot{f}_\mathrm{B}}{f_\mathrm{B}} \simeq -\frac{a_l}{c},
\label{eq:frequencies}
\end{equation}
where $a_l$ is the acceleration of the inner binary in the gravitational field of the outer star, projected along the line of sight. For the values published by \citetalias{2026NatAs..10.1020W}, we obtain, at the reference epoch:
\begin{equation}
\frac{\dot{f}}{f} = -1.5252 \times 10^{-14} \, \mathrm{s}^{-1},\;   \frac{\dot{f}_\mathrm{B}}{f_\mathrm{B}} = -1.5356(7) \times 10^{-14} \,\mathrm{s}^{-1},
\end{equation}
which are similar to within 0.7 \%\footnote{Despite the fact that the difference is small ($1.03(7) \times 10^{-16}  \,\mathrm{s}^{-1}$), it appears to be 14-$\sigma$ significant, resulting in the small gap between the spin and orbital frequency derivatives in the lower plot of Fig.~\ref{fig:periods}. This is likely a real effect, more likely a perturbation of the observed orbital period: a large intrinsic spin-down would result in a negative difference between these two terms.}. Similar equations should hold much more closely for the subsequent derivatives because, for MSPs in wide orbits, the second and higher intrinsic spin and orbital frequency derivatives are generally unmeasurably smal; or the values published by \citetalias{2026NatAs..10.1020W}, we obtain, again at the reference epoch:
\begin{eqnarray}
\frac{\ddot{f}}{f} & = -4.72 \times 10^{-23} \, \mathrm{s}^{-2},\;   \frac{\ddot{f}_\mathrm{B}}{f_\mathrm{B}} & =  -4.82(6) \times 10^{-23} \,\mathrm{s}^{-2},\\
\frac{\dddot{f}}{f} &= -2.62 \times 10^{-31} \, \mathrm{s}^{-3},\;   \frac{\dddot{f}_\mathrm{B}}{f_\mathrm{B}} & =  -2.90(7) \times 10^{-31} \,\mathrm{s}^{-3},
\end{eqnarray}
which represent differences of only 2\% and 10\%. These differences are 1.5 and 4-$\sigma$ significant, respectively; they are to be expected because, unlike for the spin frequency, no fourth and fifth derivatives of the orbital frequency were fitted. The near-identical spin and orbital frequency evolution seen in Fig.~\ref{fig:periods} is a consequence of the similarity of these values.

The implication of the close similarity of the spin and orbital frequency derivatives is clear: the large $\dot{P}$ is not intrinsic, its large value and its variation, as well as the variation of the orbital frequency, are being caused by a large, varying acceleration such as that produced by a third object in the system.

We note that \citetalias{2026NatAs..10.1020W} did consider the possibility of a hierarchical triple. However, in their treatment, they consider only circular orbits, finding that they do not provide an adequate fit to the observed data, which lead them to dismiss the hypothesis. In a circular orbit and for any positive integer, $n$, there must be a sign inversion between derivatives of order $n$ and $n+2$ \citep[in particular, see section 2.3 of][]{1997ApJ...479..948J}. For PSR~J0435+3233 all spin and orbital frequency derivatives are negative, which indeed excludes a circular orbit. However, as discussed below, it does not exclude an eccentric orbit.

From now on we will refer to the system consisting of PSR~J0435+3233 and its closer companion as the "inner" binary (and its orbit as the "inner" orbit) and the third star as the outer star. Finally, we will refer to the orbit of the inner binary around the common centre of mass with the outer star as the "outer" orbit, even though it is being traced by the inner binary.

\section{Timing PSR~J0435+3233 as a triple system}
\label{sec:timing}

\subsection{The outer orbit}

If we assume that the first observed spin derivative is much larger than the intrinsic pulsar spin-down rate, then the five spin frequency derivatives measured from the timing of PSR~J0435+3233 in \citetalias{2026NatAs..10.1020W} provide enough information for estimating the five Keplerian parameters of the outer orbit, namely the projected semi-major axis $x_\mathrm{O}$, orbital period $P_\mathrm{B,O}$, eccentricity $e_\mathrm{O}$, argument of periastron $\omega_\mathrm{O}$ and true anomaly $\nu_\mathrm{O}$ \citep{1997ApJ...479..948J,2017MNRAS.468.2114P}.

For this analysis we treat the inner binary as a point mass and hence ignore any effects that the 8-day inner orbit may have on the outer orbit. Besides the five spin frequency derivatives that \citetalias{2026NatAs..10.1020W} provide, we also include the three orbital frequency derivatives in this analysis. Because the observed first spin frequency derivative has an unknown contribution from the intrinsic spin down of the pulsar, this approach has the advantage that we can estimate the direct line-of-sight acceleration which impacts both the first derivative of the spin and orbital frequency, where we assume the latter has no or a negligible intrinsic contribution. To be able to include both the spin and orbital frequency derivatives we use the variation on the \cite{1997ApJ...479..948J} model as implemented in \citet{2016MNRAS.460.2207B}. Because of the low proper motion of the system, the Shklovskii effect \citep{1970SvA....13..562S} on the spin and orbital frequency derivatives is five orders of magnitude smaller than the values measured by \citetalias{2026NatAs..10.1020W} and can be neglected.

A likelihood function is constructed which compares the observed spin and orbit frequency derivatives from predictions by the implementation of \citet{2016MNRAS.460.2207B}. The second, up to and including the fifth, spin frequency derivatives were used, as were the three orbital frequency derivatives. The model depends on the five Keplerian parameters of the outer orbit. Given a set of orbital parameters, the model propagates the orbit from the epoch of the spin frequency derivatives (MJD\,59999) to the epoch of the orbital frequency derivatives (MJD\,59360.3145) to account for the change in motion over this time span. 

We draw random values out of uniform distributions in $\log_{10}x_\mathrm{O}$, $\log_{10} P_\mathrm{B,O}$, $e_\mathrm{O}$, $\omega_\mathrm{O}$ and $\nu_\mathrm{O}$ and use the \texttt{emcee} Markov-Chain Monte-Carlo sampler by \cite{2013PASP..125..306F} with 100 walkers and 100,000 steps to sample the posterior distributions until convergence.
The posterior distributions yield preliminary orbital solutions with $x_\mathrm{O}\approx1,900$\,s, $P_\mathrm{B, O}\approx58$\,yr, $e_\mathrm{O}\approx0.55$, $\omega_\mathrm{O}\approx34\degr$ and $\nu_\mathrm{O}\approx222\degr$. These solutions predict a periastron passage for the outer orbit in early 2036, a timeline consistent with the timing solutions discussed below.

We conclude that the observed spin and orbital frequency derivatives of PSR\,J0435+3233 can be explained by a third companion in a wide eccentric orbit around the inner binary.
\label{sec:orbit}

\subsection{Radio timing solution}

Starting from the outer orbit determined in the previous section, we then used the de-dispersed and Solar-system barycentred times-of-arrival (ToAs) published in \citetalias{2026NatAs..10.1020W} to derive a phase-coherent hierarchical triple-system timing solution for the pulsar. This was done by implementing an equivalent of the BT1P timing model of {\sc tempo} \citep{2015ascl.soft09002N} in {\sc PINT} \citep{PINTPaper1,PINTPaper2}. In this model\footnote{The PINT implementation of our hierarchical triple-system model not yet been merged into the main repository, but can be found at \url{https://github.com/vhaasteren/PINT/tree/feat/triple}.}, the light travel time delays due to the outer orbit are computed first and subtracted from the barycentric arrival times before the inner orbital phases are calculated, which accounts for the time-varying Doppler shift of the inner orbit due to the outer orbit. 

In addition to the five Keplerian parameters for the inner and outer orbits we find that a linear variation in the inner orbital period, and both linear and quadratic variations in the inner semi-major axis are necessary for an acceptable fit to the radio data. The model is a non-linear function of these parameters, and there are strong parameter correlations. This poses a problem for PINT's downhill least-squares fitting and uncertainty estimation procedures. We therefore implemented the model using JAX \citep{JAX}, which provides automatic differentiation that enables gradient-based optimisation and sampling. We used the "No U-turn Sampler" \citep{Hoffman2011+NUTS} implemented in BlackJAX \citep{blackjax} to sample the timing model parameters. For each sample, we analytically find the maximum likelihood values for the spin frequency, spin-down rate and an overall phase shift using weighted least squares, to avoid having to sample these three additional highly covariant parameters. We first ran the sampler with uniform priors on all parameters, but found that it resulted in a very large, positive frequency derivative. We therefore placed a weakly informative Gaussian prior on $\dot{f}$, centred on zero with width $5\times10^{-14}$~Hz~s$^{-1}$, larger than that of any known Galactic MSP. The posterior samples for the outer binary parameters and the spin-down rate from this procedure are shown in Fig.~\ref{fig:corner}, with the full posterior distribution given in Appendix~\ref{app:results}, Fig.~\ref{fig:full_corner}. The best-fitting parameters and their uncertainties are reported in Table~\ref{tab:solution}.

\begin{figure*}
 \sidecaption
    \includegraphics[width=12cm]{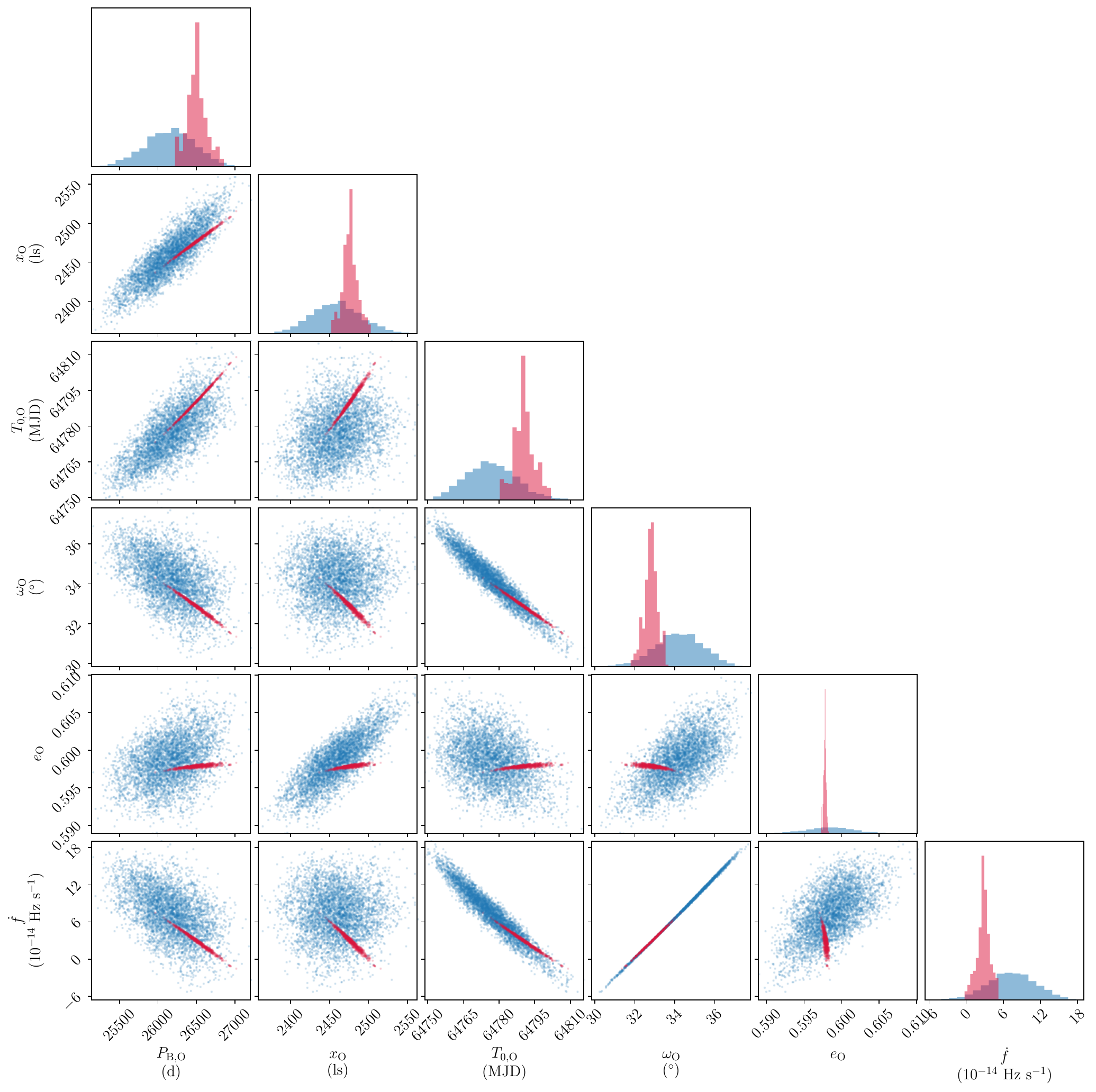}
     \caption{Corner plot showing the Keplerian parameters for the outer orbit and the spin frequency derivative obtained from our timing analysis. Blue points show samples from radio-only timing after thinning the sample chains by a factor of 100. The red points (and marginal distributions on the diagonal) contain 99.9\% of the $\gamma$-ray re-weighted posterior probability, which results in a significant narrowing of the uncertainties of some timing parameters. A similar plot showing all model parameters is provided in Fig.~\ref{fig:full_corner}.}
     \label{fig:corner}
\end{figure*}

\begin{table*}
\centering
\caption{Timing solution for the PSR~J0435+3233 triple system.\label{tab:solution}}
\begin{tabular}{lcc}
\hline
\hline
  \multicolumn{3}{c}{Data properties} \\
\hline
Solar system ephemeris     \dotfill & \multicolumn{2}{c}{DE438} \\
Time units                 \dotfill & \multicolumn{2}{c}{TDB} \\
First radio ToA (MJD)      \dotfill & \multicolumn{2}{c}{59101.877} \\
Last radio ToA (MJD)       \dotfill & \multicolumn{2}{c}{60897.993} \\
Reference epoch            \dotfill & \multicolumn{2}{c}{59999} \\
Number of radio ToAs       \dotfill & \multicolumn{2}{c}{997} \\
First $\gamma$-ray photon arrival (MJD) \dotfill & \multicolumn{2}{c}{54683} \\
Last $\gamma$-ray photon arrival (MJD) \dotfill & \multicolumn{2}{c}{61135} \\

\hline
 \multicolumn{3}{c}{Fixed parameters} \\
\hline
Right ascension, $\alpha$ (hh:mm:ss.s)          \dotfill & \multicolumn{2}{c}{04:35:33.760884} \\
Declination, $\delta$  ($\circ,\arcmin,\farcs$) \dotfill & \multicolumn{2}{c}{$+$32:33:07.93460} \\
Proper motion in $\alpha$, $\mu_\alpha$         \dotfill & \multicolumn{2}{c}{1.36} \\
Proper motion in $\delta$, $\mu_\delta$         \dotfill & \multicolumn{2}{c}{$-$2.2} \\
DM ($\rm cm^{-3} \, pc$)        \dotfill  &    \multicolumn{2}{c}{37.346} \\
\hline
\hline
\multicolumn{3}{c}{Measured quantities} \\
\hline
& Radio-only & Joint radio/$\gamma$-ray \\  
\hline
\multicolumn{3}{c}{Spin parameters} \\
\hline
Spin frequency, $f$ ($\rm Hz$)                  \dotfill & 312.71122(4) & 312.711168(4) \\
Derivative of $f$, $\dot{f}$ ($10^{-15}$ Hz s$^{-1}$) \dotfill & 110(40) & 30(10) \\
\hline
 \multicolumn{3}{c}{Inner orbit} \\
\hline
Binary model                \dotfill & \multicolumn{2}{c}{DD} \\
Orbital period, $P_\mathrm{B}$ (d)         \dotfill &        7.9981454(9) & 7.99814665(8) \\
Orbital period derivative, $\dot{P}_\mathrm{B}$ ($10^{-12}$)     \dotfill &   $-140$(90) & 36(24)\\
Projected semi-major axis, $x$ (s)    \dotfill &            7.9781037(9) & 7.9781051(2)\\
Derivative of proj. semi-major axis, $\dot{x}$ ($10^{-15} $)      \dotfill &   362(5) & 365(5)    \\
Second derivative of proj. semi-major axis, $\ddot{x}$ ($10^{-21}$ s$^{-1}$)      \dotfill &   2.00(8) & 1.90(9)  \\
Eccentricity, $e$                       \dotfill &       0.00016279(2) & 0.00016278(2) \\
Periastron passage, $T_0$ (MJD)    \dotfill &       59360.3470(4)  & 59360.3467(1)\\
Longitude of periastron, $\omega$ ($\deg$)         \dotfill &        121.112(6) & 121.110(6)\\
\hline
 \multicolumn{3}{c}{Outer orbit} \\
\hline
Binary model                \dotfill & \multicolumn{2}{c}{DD} \\
Orbital period, $P_\mathrm{B, O}$ (d)       \dotfill &        26040(320) & 26500(120) \\
Projected semi-major axis, $x_\mathrm{O}$ (s)        \dotfill &           2460(30) & 2477(9)\\
Eccentricity, $e_\mathrm{O}$       \dotfill &           0.600(3) & 0.5978(2)\\
Periastron passage, $T_\mathrm{0, O}$ (MJD)        \dotfill &       64769(11)& 64791(4) \\
Longitude of periastron, $\omega_\mathrm{O}$  ($\deg$)        \dotfill &        35.2(1.4) & 32.8(3) \\
\hline
\hline
\multicolumn{3}{c}{Solution properties}\\
\hline
$\chi^2$ of the radio fit        \dotfill & 995.9 & 1004.5\\
Reduced $\chi^2$ of radio ToAs           \dotfill & 1.002 & 1.01\\
Residual rms ($\upmu$s) of radio ToAs    \dotfill & 1.49 & 1.50\\
$\gamma$-ray weighted $H$-statistic \dotfill & 112.5 & 171.9\\
\hline\hline
\multicolumn{3}{c}{Derived properties}\\
\hline
Spin period, $P$ (ms) \dotfill & 3.1978386(4) & 3.19783910(4) \\
Spin-down rate, $\dot{P}$ ($10^{-18}$) \dotfill & $-1.1(4)$ & $-0.3(1)$ \\
\hline
\end{tabular}
\tablefoot{The parameters presented without uncertainties are adopted from \citetalias{2026NatAs..10.1020W}; the parameters with uncertainties are derived from our fit of the timing data; the uncertainties are 68\% confidence limits. The timing solution on the left is the mean of the posterior samples from timing using only the barycentric ToAs published in \citetalias{2026NatAs..10.1020W}. The solution on the right has the maximum joint radio/$\gamma$-ray log-likelihood, with uncertainties derived from the $\gamma$-ray re-weighted posterior samples.}
\end{table*}

\begin{figure}
    \centering
    {\includegraphics[width=\columnwidth]{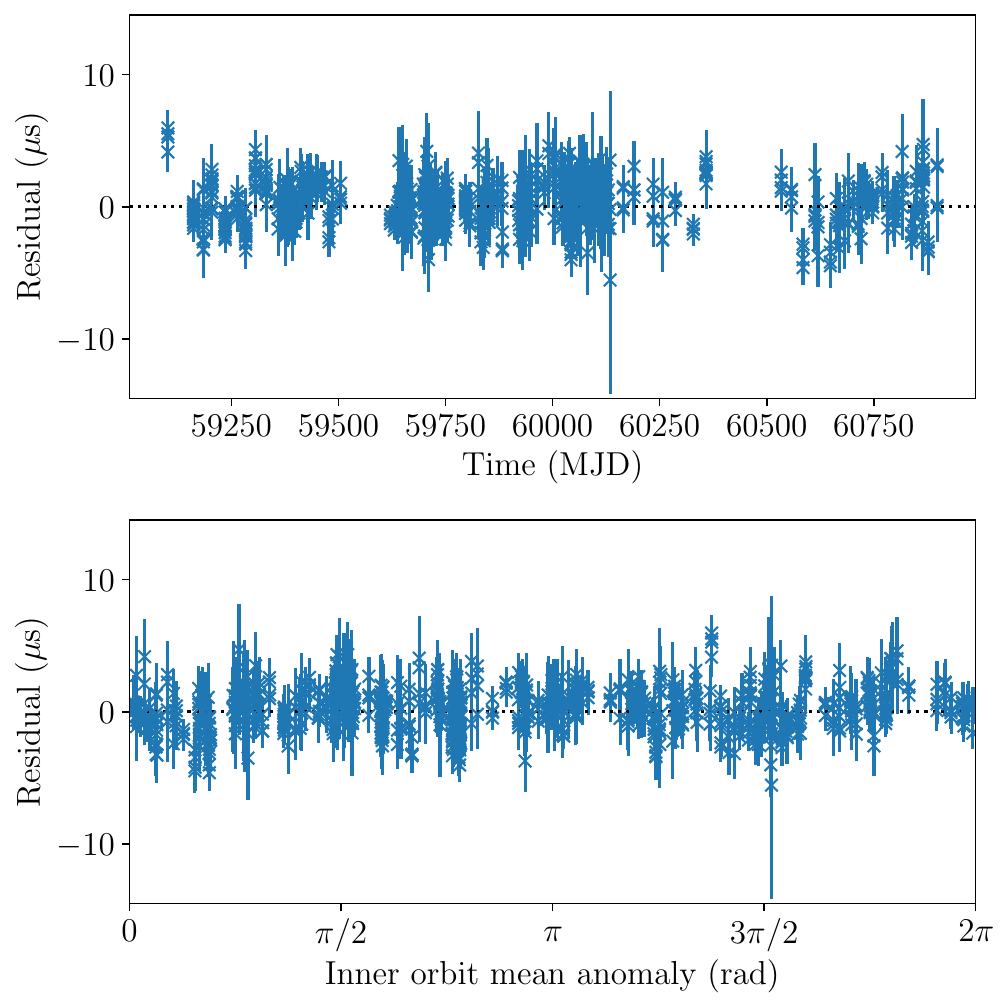}}
     \caption{Timing residuals obtained from the FAST barycentric ToAs of \citetalias{2026NatAs..10.1020W} using the solution in the second column of Table~\ref{tab:solution}. The upper panel shows residuals as a function of time; the lower panel shows these as a function of the mean anomaly (orbital phase) of the inner orbit.}
     \label{fig:residuals}
\end{figure}

\subsection{Radio plus $\gamma$-ray timing solution}
\label{sec:gamma}

We used the {\it Fermi}-LAT data to further test this triple-system model. In contrast to the Taylor series model used by \citet{2026arXiv260716119Z}, if the triple-system model is correct then it should predict the pulsar spin phase over a longer time interval, ideally back to the start of the {\it Fermi}-LAT data in 2008. 

We used \texttt{SOURCE}-class photons according to the \texttt{P8R3\_SOURCE\_V3} instrument response functions \citep{Pass8,Bruel2018+P305}, using the same angular and energy cuts used in \cite{4FGL}, and computed photon probability weights \citep{Kerr2011} according to the 4FGL-DR4 \citep{4FGLDR3, 4FGLDR4} point-source spectral and spatial models, and the associated rescaled \texttt{gll\_iem\_v07} interstellar emission and \texttt{iso\_P8R3\_SOURCE\_V3\_v1.txt} isotropic background models\footnote{\url{https://fermi.gsfc.nasa.gov/ssc/data/access/lat/BackgroundModels.html}}. Folding these photons using the best-fitting radio timing solution allows a detection of the $\gamma$-ray pulsations as far back as MJD 57500 (late 2014), i.e., up to 6 years earlier than in the detection of \cite{2026arXiv260716119Z}. Still, even this improved ephemeris is not precise enough to accurately predict the spin phase before that time. 

To further refine the model, we folded the {\it Fermi}-LAT data using all posterior samples from our radio timing and computed the weighted $H$ statistic \citep{deJager1989,Kerr2011} to evaluate the pulsation significance. We find a cluster of samples whose $\gamma$-ray $H$-statistic is significantly higher than the $H \approx 66$ obtained by \citet{2026arXiv260716119Z}. The sample with the highest $H$-statistic value has $H = 172$, and appears to result in a phase-coherent solution reaching back to the start of the LAT data in August 2008. 

\begin{figure}
    \centering
    {\includegraphics[width=\columnwidth]{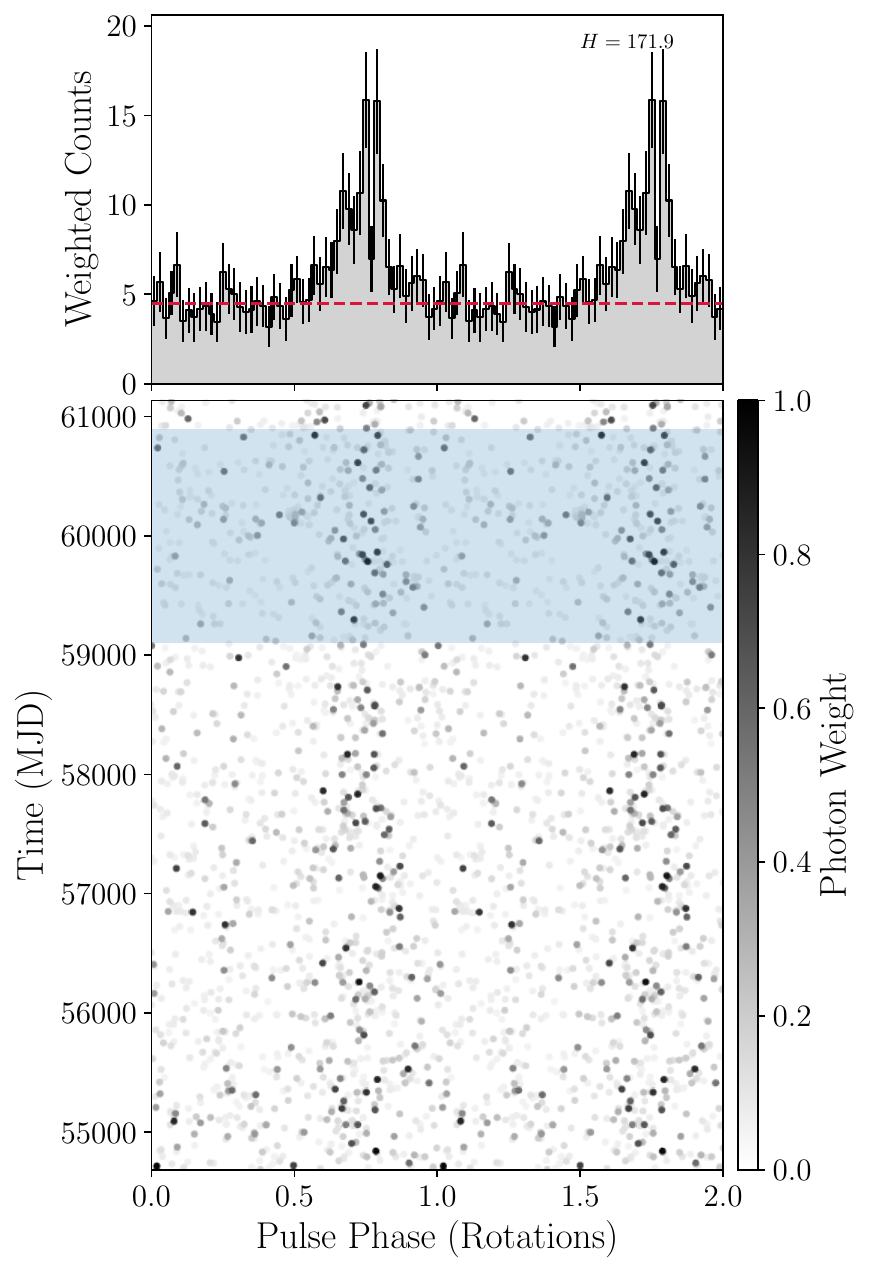}}
     \caption{Detection of $\gamma$-ray pulsations for PSR~J0435+3233. The spin phases of the $\gamma$-ray photons, shown as points in the lower panel, were calculated using the timing solution in the second column of Table~\ref{tab:solution}, which maximises the joint radio and $\gamma$-ray pulsation log-likelihood. Phase zero occurs at the (arbitrary) barycentric reference epoch of MJD 59999. The blue shaded region indicates the time interval spanned by FAST observations. The top panel shows in the integrated pulse profile, with the background level (estimated from the photon weights as in \citealt{2023ApJ...958..191S}, Section 5.1) shown by the red dashed line.}
     \label{fig:gamma}
\end{figure}

Existing software for jointly fitting radio and gamma-ray timing data \citep[e.g.,][]{Burgay2024,Clark2026+shoogle} requires a model that is approximately linear in the parameters, which our triple-system model is not. To obtain a refined estimate of the model parameters incorporating this $\gamma$-ray detection, we therefore adopted the technique of importance sampling \citep{vanDijkKloek1978+ImportanceSampling}. We took all posterior samples from the radio timing and re-weighted them by the inverse of the $H$-statistic false-alarm probability, $W \propto \exp(0.398405 H)$ \citep{Kerr2011}, as a proxy for a $\gamma$-ray pulsation log-likelihood that is agnostic about the pulse shape. The parameters with the highest combined radio and $\gamma$-ray log-likelihood and uncertainties from these re-weighted samples are reported in Table~\ref{tab:solution}, with the resulting radio ToA residuals and $\gamma$-ray photon phases shown in Figs.~\ref{fig:residuals} and ~\ref{fig:gamma}, respectively.

Despite the relatively low precision of the {\em Fermi}-LAT timing data compared to the FAST ToAs, the $\gamma$-ray detection results in a significantly higher precision for the timing parameters, especially those related to the outer orbit; this is a direct result of the increased timing baseline added by the $\gamma$-ray pulsations, which is now 18 years, or about 25\% of the full orbit.

\section{The outer star}
\label{sec:third_star}
Optical and (near-)infrared catalogs report a counterpart within $\sim20$\,mas from the celestial position of PSR\,J0435+3233. We will refer to this object by its identification in the 2MASS catalog \citep{2006AJ....131.1163S}: 2MASS\,J04353375+3233080. The \textit{Gaia} DR3 astrometric solution \citep{2023A&A...674A...1G} of 2MASS\,J04353375+3233080 reports a proper motion of $\mu=3.60(6)$\,mas\,yr$^{-1}$ with a position angle of $\theta=168.9(1.3)$\,\degr, which is comparable to that of the pulsar, for which \citetalias{2026NatAs..10.1020W} report $\mu=2.58(17)$\,mas\,yr$^{-1}$ at $\theta=148(2)$\,\degr. Propagating the \textit{Gaia} astrometry to the epoch of the pulsar ephemeris (MJD\,59999; 2023 February 24) yields a separation of 11.0\,mas. This is near the minimum separation that the motion of PSR\,J0435+3233 and 2MASS\,J04353375+3233080 reached on the sky, which occurred on 2023 April 23.

Within 30\,\arcmin\ of the location of PSR\,J0435+3233, the \textit{Gaia} DR3 catalogue has a stellar density of 5.93\,arcmin$^{-2}$. As a result, the probability of finding an unrelated star within 11\,mas of the position of PSR\,J0435+3233 by chance is $6.2\times10^{-7}$. This probability decreases to $6.8\times10^{-8}$ for \textit{Gaia} DR3 catalogued objects that are brighter than 2MASS\,J04353375+3233080 (\textit{Gaia} magnitude $G<16.66$). 

\begin{figure}
    \centering
    {\includegraphics[width=\columnwidth]{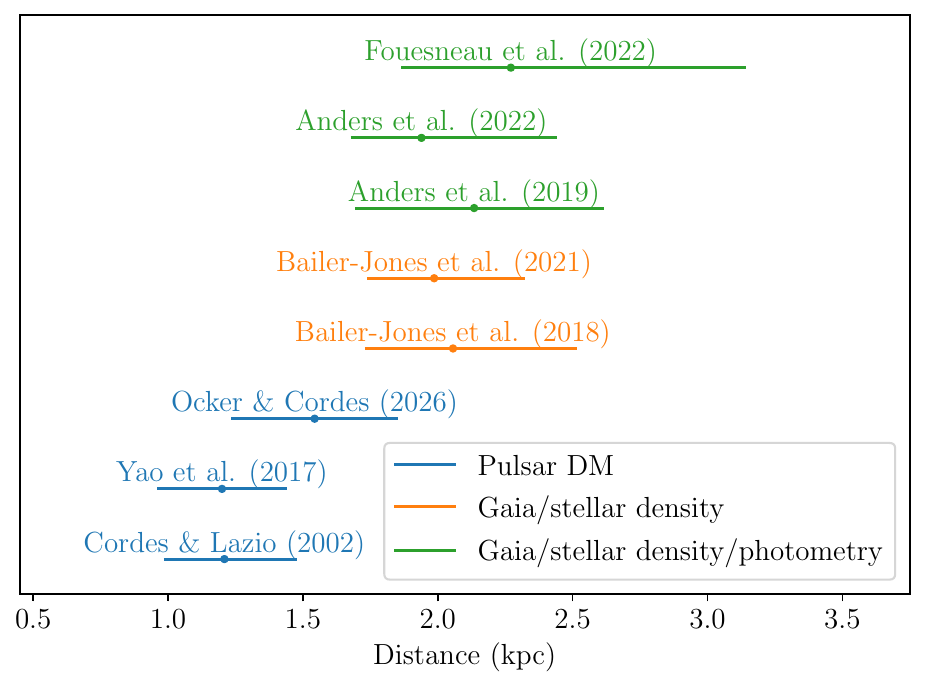}}
     \caption{Distance estimates of PSR\,J0435+3233 based on the pulsar dispersion measure and Galactic electron density models, and of the outer companion 2MASS\,J04353375+3233080 based on \textit{Gaia} parallax, Galactic stellar density models and/or broadband optical and near-infrared photometry.}
      \label{fig:distances}
\end{figure}

Distance estimates based on the \textit{Gaia} parallax of 2MASS\,J04353375+3233080 are available using different methods. \citet{2018AJ....156...58B,2021AJ....161..147B} use the \textit{Gaia} DR2 and EDR3 parallax, respectively, combined with priors from Galactic stellar density models to invert the parallax, while \citet{2019A&A...628A..94A,2022A&A...658A..91A,2022A&A...662A.125F} further include broadband optical and near-infrared photometry combined with predictions from stellar models to obtain photo-astrometric distances. These models place 2MASS\,J04353375+3233080 at a distance of around 2.0 to 2.2\,kpc. This is somewhat larger but still consistent within the uncertainties with the distances estimated from models of the Galactic distribution of free electrons, see Fig.\,\ref{fig:distances}. The distances derived from the NE2001 \citep{2002astro.ph..7156C}, YMW16 \citep{2017ApJ...835...29Y} and NE2025 \citep{2026ApJ..1002....3O} models respectively, are 1.2, 1.2 and 1.5 kpc (using implementations of these models in {\sc PyGEDM} by \citealt{2021PASA...38...38P}). 

Based on the positional coincidence, the low probability of a chance alignment and the consistent distances for PSR\,J0435+3233 and 2MASS\,J04353375+3233080, we conclude that the latter is the tertiary companion to the pulsar. 

The photo-astrometric distance models also provide stellar parameters of 2MASS\,J04353375+3233080 and identify it as a main-sequence star. \citet{2019A&A...628A..94A} and \citet{2022A&A...658A..91A} find effective temperatures in the range of $T_\mathrm{eff}=5900$ to $7300$\,K, surface gravities of $\log g=4.1$ to 4.4\,g\,cm$^{-2}$ and masses of $M=1.0$ to 1.4\,M$_\odot$. The model by \citet{2022A&A...662A.125F} predicts comparable properties ($T_\mathrm{eff}=6100$ to $8200$\,K, $\log g=4.1$ to $4.4$\,g\,cm$^{-2}$ and $M=1.1$ to 1.7\,M$_\odot$). The outer companion would therefore be classified with a spectral type in the range of G0V and F0V, possibly extending as early as A5V following the \citet{2022A&A...662A.125F} model.

We note that at a distance of 2\,kpc, a $\gtrsim 0.3\,\rm M_\odot$ white dwarf in the inner binary would have a {\it Gaia} $G$-band magnitude $G>22.8$ for $T_\mathrm{eff}<30000$\,K\footnote{\url{http://www.astro.umontreal.ca/~bergeron/CoolingModels}} \citep{1995PASP..107.1047B} and would not be detectable with \textit{Gaia}. The 11\,mas separation to 2MASS\,J04353375+3233080 with $G=16.66$ would also greatly hinder detecting the white dwarf.

\begin{figure*}
     \sidecaption
    {\includegraphics[width=12cm]{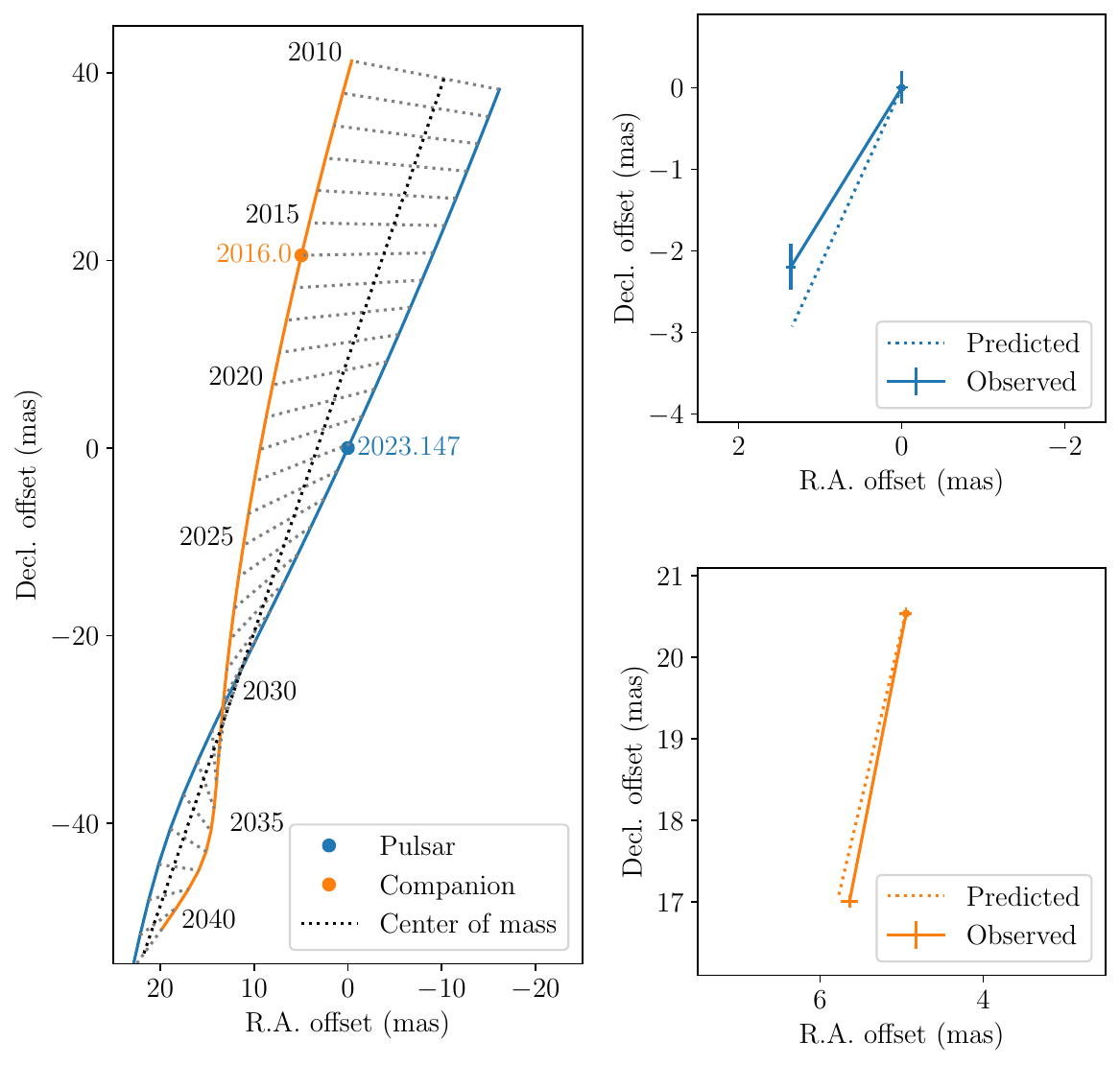}}
     \caption{{\it Left}: the motion of the inner binary pulsar, PSR~J0435+3233 (blue), 2MASS\,J04353375+3233080 (orange) and the centre of mass of the system (black dotted line) projected on the plane of the sky for a 30-yr timespan using the orbital parameters of the outer orbit from Table\,\ref{tab:solution}. Positions are plotted relative to the position of the pulsar at the epoch of the \citetalias{2026NatAs..10.1020W} timing ephemeris (2023.147). The measured position of 2MASS\,J04353375+3233080 is given by the \textit{Gaia} DR3 astrometric solution at the \textit{Gaia} epoch of J2016.0. The lighter dotted lines link the position of the inner binary and 2MASS\,J04353375+3233080 at the start of every year in the 30-year time span.
     {\it Right:} The two insets show the motion over 1\,yr as predicted by the orbital solution as well as the proper motions observed for the inner binary (from timing, top) and for 2MASS\,J04353375+3233080 (from GAIA, bottom). These plots neglect the $\sim0.5$\,mas yearly variations due to parallax.}
      \label{fig:projected_orbit}
\end{figure*}

Assuming masses for the inner binary and the outer companion of $M_\mathrm{b} = 2\, \rm M_\odot$ and 
$M_\mathrm{O} = 1.2\, \rm M_\odot$, respectively, the semi-major axis of the outer orbit determined from pulsar timing (Table~\ref{tab:solution}) is constrained to $a=25.5$\,AU, which in turn sets $\sin i_\mathrm{O} = 0.516$. This implies an orbital inclination of the outer orbit of $i_\mathrm{O}=31\degr$ or $149\degr$\footnote{Assuming for the inner orbit $i \sim i_\mathrm{O}$, and $M_\mathrm{p} = 1.4 \, \rm M_\odot$, we obtain an inner companion mass $M_\mathrm{i} \sim 0.6 \, \rm M_\odot$, and thus an inner binary mass of $M_\mathrm{b} \sim 2.0 \, \rm M_\odot$, which is consistent with the value used in the derivation of $i_\mathrm{O}$.}. Finally, we can project an orbit with these parameters onto the sky and compare the predicted binary motion to the measure position and proper motion of 2MASS\,J04353375+3233080 relative to PSR\,J0435+3233. Here, we implicitly assume that both the \textit{Gaia} astrometry of 2MASS\,J04353375+3233080 and the pulsar timing astrometry of PSR\,J0435+3233 at their respective epochs are not affected by higher order effects of the binary orbit. Under this assumption, this projection constrains the position and proper motion of the centre of mass of the outer binary, as well as the distance $d$ and orientation of the outer orbit on the sky (via the right ascension of the ascending node $\Omega$). We find a solution that has $i=149\degr$ and yields $d=2.2$\,kpc and $\Omega=56\degr$. The motion of PSR\,J0435+3233 and 2MASS\,J04353375+3233080 on the sky is shown in Fig.\,\ref{fig:projected_orbit}.

\section{Discussion}
\label{sec:discussion}

We have demonstrated that the exceptionally high spin-down rate of PSR\,J0435+3233 identified by \citetalias{2026NatAs..10.1020W} can be explained by unmodelled line-of-sight acceleration due to a third companion in a wide eccentric orbit. 

We derived and fitted a triple-system model that accurately describes the radio pulse arrival times detected by FAST and the $\gamma$-ray photon arrival times measured by the {\it Fermi} LAT over a significantly longer timing baseline.
Furthermore, the orbital configuration derived from this solution can account for the separation between PSR~J0435+3233 and the outer star, 2MASS\,J04353375+3233080 and can also explain their small proper motion difference.
We now make a few general remarks about this solution. 

The residual RMS of the FAST ToAs, $1.49 \, \upmu \rm s$, is significantly lower than the residual RMS reported by \citetalias{2026NatAs..10.1020W}, $1.79 \, \upmu \rm s$. In particular, the reduced $\chi^2$ of 1.002, obtained without any adjustments of the ToA uncertainties would suggest a very good quality of fit. However, an analysis of Fig.~\ref{fig:residuals} suggests that there might still be some trends in the data. This is expected given the complexity of timing a pulsar in a triple star system, which will likely require numerical integration of the inner and outer orbit, as in the case of PSR~J0337+1715 \citep{2014Natur.505..520R,2018Natur.559...73A,2020A&A...638A..24V}, to fully account for orbital perturbations, tidal effects and time dilation, effects which are not included in our simplified model consisting of two non-interacting Keplerian orbits.

In the timing ephemeris we cannot fit for the DM and its time derivatives, nor for position and proper motion because the set of ToAs published in \citetalias{2026NatAs..10.1020W} have already been barycentered and de-dispersed. This also means that the uncertainties in our parameters will be slightly under-estimated because the covariances between the timing parameters and the DM, position and proper motion cannot be taken into account. Furthermore, as shown in Fig.~\ref{fig:projected_orbit}, the apparent motion of the pulsar over long time spans will not follow a constant proper motion; our model does not yet account for this effect, which could be significant over the {\it Fermi}-LAT data span.

Most of the spin and orbital frequency derivatives in the solution of \citetalias{2026NatAs..10.1020W} are absent in the timing model given in Table\,\ref{tab:solution}, as these effects are now accounted for by the R\o{}mer delay of the outer orbit. However, the intrinsic spin-frequency derivative remains highly uncertain, as its value is highly covariant with the outer orbital parameters and the derivative of the inner orbital period ($\dot{P}_{\mathrm{B}}$). From the $\gamma$-ray re-weighted posterior samples, we estimate $\dot{f} = (2.8 \pm 1.0)\times 10^{-14}$~Hz~s$^{-1}$ ($\dot{P} = (-2.8 \pm 1.0)\times10^{-19}$). This slight preference towards an (unphysical) positive spin-frequency derivative may be a further sign that a full three-body model will be required here.

Although we cannot yet measure $\dot{P}$ precisely, it is clear from its uncertainty that it is at least two orders of magnitude smaller than the value claimed by \citetalias{2026NatAs..10.1020W}. This has several astrophysical implications. The first one is that we cannot yet constrain the efficiency of the $\gamma$-ray emission, although we can safely say that it is not likely to be anomalously low, as suggested by \cite{2026arXiv260716119Z}. Second, as noted by \citet{2026arXiv260718219M}, the much smaller intrinsic $\dot{P}$ means that the non-detection of continuous gravitational wave emission from the pulsar no longer constrains the fraction of spin-down power emitted as gravitational waves.

The parameters derived from the initial spin and orbital frequency derivatives are somewhat different from those derived in the timing analyses, with the latter showing that the outer orbit has a larger period (72 versus 58 yr) and eccentricity (0.60 versus 0.55).
The cause of this is unclear, however, the high significance of the spin frequency derivatives (even the fifth) published by \citetalias{2026NatAs..10.1020W} is such that additional spin frequency derivatives are likely to be detectable in their data. If they were fitted, they would certainly alter (and degrade the uncertainty of) the first five derivatives. Such information cannot be taken into account in an analysis of those derivatives, but is always fully used in a timing analysis. Furthermore, the $\gamma$-ray data favours, and extends, the results from the radio timing analysis.

Part of the difference is that our timing model also includes significant contributions to $\dot{P}_\mathrm{B}$ and $\dot{x}$ that cannot be attributed to the acceleration of the inner binary around the common centre of mass. 
Our estimate of $\dot{x}$, $3.65(5) \times 10^{-13}$, is of a similar magnitude to that measured by
\citetalias{2026NatAs..10.1020W}, $\dot{x} = 5.95(2) \times 10^{-13}$. Part of the reason for the numerical difference in the values of $\dot{x}$ is that, in addition, we fit for $\ddot{x}$, which significantly reduces the rms uncertainty and yields a highly significant estimate of this quantity.
A possible contribution from a secular change in the apparent orbital inclination caused by the proper motion \citep{1996ASPC..105..525A,1996ApJ...467L..93K} is not only too small to account for the value of $\dot{x}$, but cannot certainly account for $\ddot{x}$.
We find it very likely that these derivatives result instead from the unusual variation of the relativistic time dilation of the pulsar and from tidal effects that result in mutual interactions between the inner and outer orbits \citep{1962P&SS....9..719L,2025A&A...697A.166D}, as observed in PSR~J0337+1715 \citep{2014Natur.505..520R}.
A study of these effects will be key to a full characterisation of the system and for lifting degeneracies between orbital parameters, particularly in a system such as PSR~J0435+3233 which is weakly relativistic.

We can estimate the magnitude of such mutual interaction effects by using the perturbative model presented in appendix B of \citet{2025A&A...693A.143V}. This model stipulates that the amplitude of the timing delay caused by the secular deformation of the inner binary by the outer star increases linearly with time at a rate (Eq. B.23 of \citealt{2025A&A...693A.143V}),
\begin{equation}
 \rho = \frac{GM_{\rm O} x P_\mathrm{B}}{2\pi c^3 x_\mathrm{O}^3}  \frac{M_\mathrm{O}^3}{(M_\mathrm{b} + M_\mathrm{O})^3}\frac{M_\mathrm{b}}{M_\mathrm{i}}\frac{\sin^3 i_\mathrm{O}}{\sin i} \sim 5\times 10^{-12},
\end{equation}
where we assumed as above $M_\mathrm{b} = 2 \, \rm M_\odot$, $M_\mathrm{O} = 1.2\, \rm M_\odot$, coplanar orbits of inclinations $i \sim i_\mathrm{O} \sim 31^\circ$, an inner companion mass of $M_\mathrm{i} \sim 0.6\, \rm M_\odot$, and the values of Table \ref{tab:solution} for the other parameters. Over the radio data span of $T_\mathrm{span}\simeq 1800\;\mathrm{d}$, this leads to a maximum amplitude for the extra delay of $\Delta \sim  \rho T_\mathrm{span}  \sim 0.8 \;\mathrm{ms}$. Although this delay can be absorbed into other parameters, it is substantial and thus potentially detectable and thus promising for the future characterisation of the system.

\section{Summary and prospects}
\label{sec:conclusions}

\begin{figure}
    \centering
    {\includegraphics[width=1.05\columnwidth]{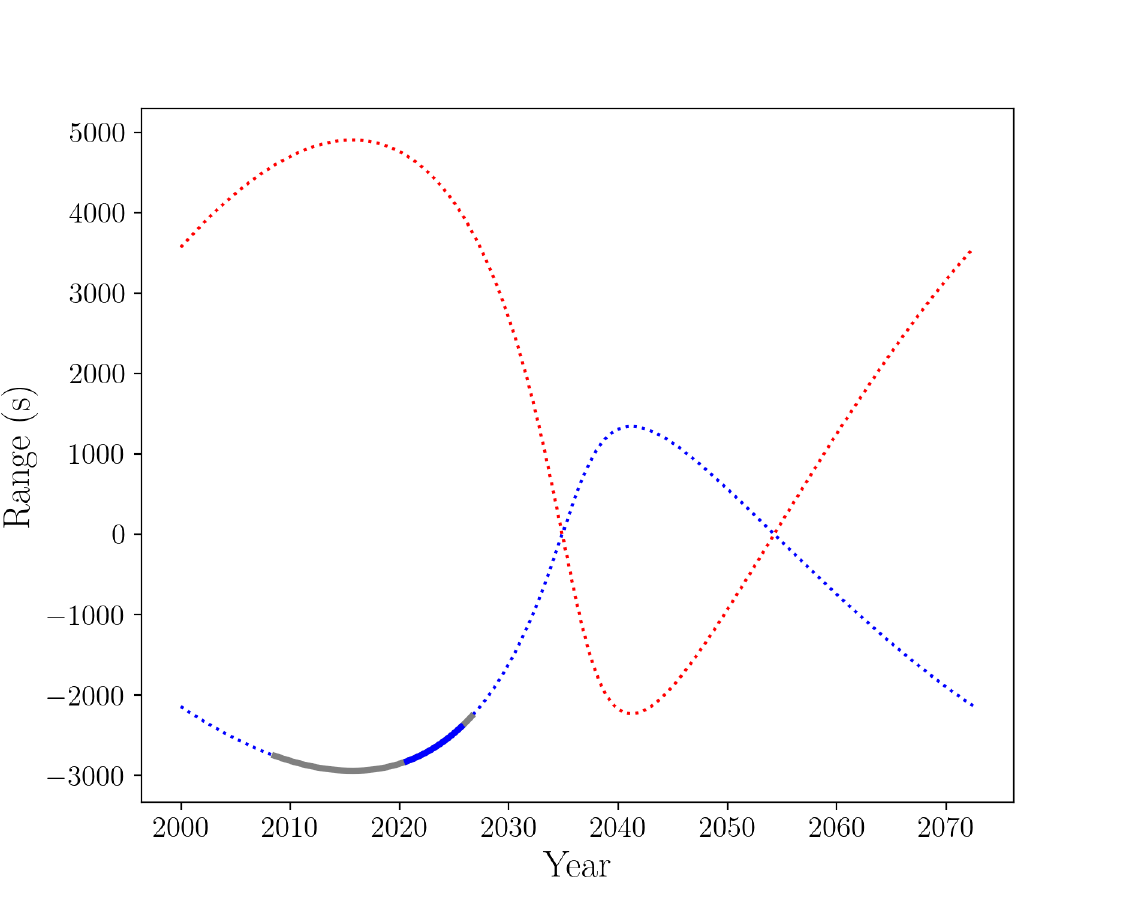}}
    {\includegraphics[width=1.05\columnwidth]{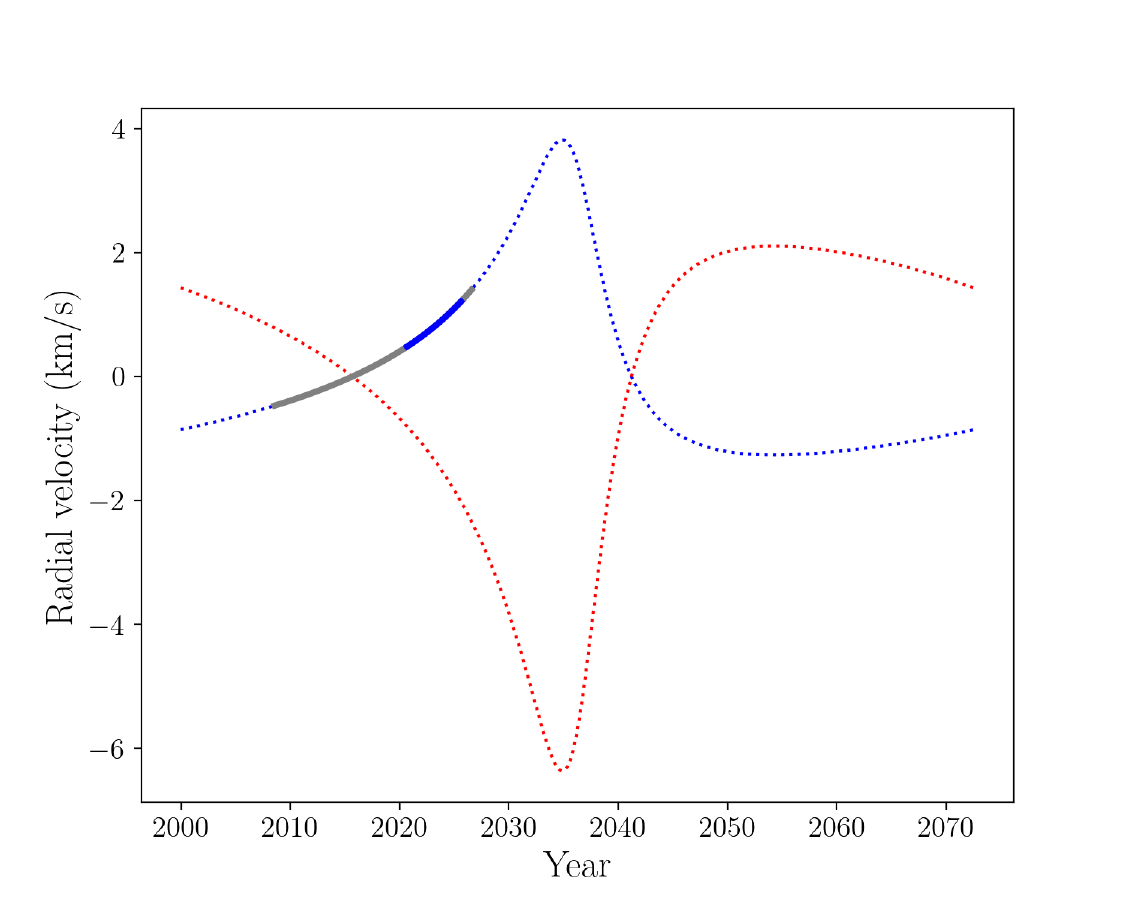}}
     \caption{Evolution of the range, or R\o{}mer delay (the light-travel-time to the Solar System Barycentre minus the same quantity for the centre of mass of the triple, top panel) and radial velocity (minus the radial velocity of the centre of mass of the triple, bottom panel) for the outer star (red) and the centre of mass of the inner binary (blue). The time axis covers one full orbit and is centred at periastron. The range and orbital radial velocity of the inner binary were calculated from the relevant parameters in the second column of Table~\ref{tab:solution}: the gray lines indicate the {\em Fermi} timing baseline, the solid blue lines indicates the radio timing baseline of \citetalias{2026NatAs..10.1020W}, the dotted blue lines represent the prediction of the ephemeris. For the motion of the outer star (dotted red lines) we assume, in addition, an outer mass ratio $R_\mathrm{O} = M_\mathrm{O}/M_\mathrm{b} = 1.2 / 2.0 = 0.6$; this quantity might soon be constrained by RV measurements of the outer star.
     }
     \label{fig:range_velocity}
\end{figure}

In this work, we conclude that PSR~J0435+3233 is a member of a triple star system, with a typical MSP--white-dwarf binary and a stellar-mass main-sequence star orbiting a common centre of mass. This naturally explains not only the large observed $\dot{P}$, but also its $\sim$ 60\% increase during the time of the observations reported in \citetalias{2026NatAs..10.1020W} and, more importantly, the very similar (but not identical) evolution of the orbital and spin frequency derivatives. From the spin and orbital frequency derivatives, we estimate the orbital parameters of the outer orbit, which is rather wide and eccentric.

Using this model as a starting point we fit a triple-system model to the FAST ToAs, and refine this by testing model parameters for $\gamma$-ray pulsations in {\it Fermi}-LAT observations, eventually finding a model that includes an outer orbit with $P_\mathrm{B,O}=72$\,yr and $e_\mathrm{O}=0.60$ which recovers $\gamma$-ray pulsations throughout the entire {\it Fermi}-LAT observing timespan. Despite the much lower instantaneous precision of the $\gamma$-ray timing compared to radio timing, the increased coverage of the outer orbit from $\sim$7\% to $\sim$25\% of the orbital period (see Figure~\ref{fig:range_velocity}) results in significantly higher precision of the parameters for the outer orbit. Finally, we identify the outer star of the system in the {\it Gaia} catalogue, 2MASS\,J04353375+3233080, that appears to be an unevolved main sequence star with a mass between 1.0 and $1.7\, \rm M_{\odot}$. The derived outer orbit is consistent with the small observed differences in position and proper motion of PSR~J0435+3233 and 2MASS\,J04353375+3233080.

Our timing solution includes significant measurements of $\dot{x}$, $\ddot{x}$ and $\dot{P}_\mathrm{B}$; these are probably describing tidal effects in the inner orbit and the varying relativistic time dilation of the pulsar relative to our reference frame; a fully self-consistent modelling including these effects will likely provide additional information on the masses, distance and orbital orientation of the system. 

The prospects for the continued study of this system are very enticing. First, it is clear that the mere extension of the timing baseline will quickly improve the precision of the outer orbital parameters. This improvement will accelerate over the next 10 years as we approach periastron in 2036, where the change in the range (commonly known as the R\o{}mer delay) of the centre of mass of the inner binary will be very steep and non-linear (see top plot of Fig.~\ref{fig:range_velocity}). Second, the {\it Gaia} DR4 data release will include measurements of $\alpha$ and $\delta$ for several distinct epochs, enabling a measurement of any non-linearities in the proper motion of the outer star. Similar measurements of the transversal motion of the pulsar will be detectable as a non-linear variation of the coordinates of the pulsar, further solidifying our estimates of $i$ and $\Omega$ of the outer orbit and of the distance to the system.

More importantly, with the current orbital model and assumed mass ratio, we predict (assuming an "outer" mass ratio $R_\mathrm{O} = M_\mathrm{O}/M_\mathrm{b} = 1.2 / 2.0 = 0.6$) a significant ($\sim 4 \, \rm km \, s^{-1}$) decrease in the radial velocity of the outer star until 2036, followed by a $\sim 8 \, \rm km \, s^{-1}$ increase until 2050 (see lower plot in Fig.~\ref{fig:range_velocity}). Measuring this variation precisely via high-resolution optical spectroscopy will likely be possible given its brightness. Such measurements will yield precise estimates of the systemic radial velocity and, together with the proper motion, will enable a determination of the 3-D velocity of the system. Furthermore, they will also allow a precise, theory-independent determination of $R_\mathrm{O}$. High resolution optical spectroscopy will also improve the stellar parameters of the outer star, including its mass, age and the system's distance.

What makes these measurements especially interesting is the fact that timing the first discovered triple system in the Galaxy, PSR~J0337+1715 \citep{2014Natur.505..520R} has allowed a test of the universality of free fall (UFF) for self-gravitating objects of unprecedented precision: the accelerations of the pulsar and inner white dwarf (WD) in the field of the outer WD have a fractional difference $|\Delta| < 2.6 \times 10^{-6}$ \citep{2018Natur.559...73A}. This important result was later confirmed and improved upon using a completely different data set from a different observatory and using completely independent data analysis algorithms by \cite{2020A&A...638A..24V,2025A&A...693A.143V}, who find that $|\Delta| < 2.0 \times 10^{-6}$. These limits represent the most stringent constraints on some alternative theories of gravity: as an example, from the Jordan-Fierz-Brans-Dicke (JFBD) gravity theory, this yields a conservative 95\% lower limit on the Brans-Dicke parameter\footnote{This is a dimensionless parameter that determines the strength of the coupling between a scalar field and the spacetime geometry in JFBD gravity \citep{1961PhRv..124..925B}. As its approaches infinity, the theory converges asymptotically to general relativity.} of $\omega_\mathrm{BD} > 150,000$ \citep{2024LRR....27....5F}, which is significantly more constraining than the best previous test obtained from the tracking of the Cassini mission, which had obtained $\omega_\mathrm{BD} > 40,000$ \citep{2003Natur.425..374B}. The importance of such tests of the UFF for self-gravitating objects results from the fact that they are conceptually the simplest, most direct and precise tests of the strong equivalence principle, a fundamental prediction of general relativity that distinguishes it from most other viable alternative gravity theories \citep{2018tegp.book.....W}. 

In the case of a violation of the UFF, there is a polarisation of the inner orbit that is detectable as a "forced" eccentricity vector in the direction of the outer mass \citep{1968PhRv..170.1186N,1991PhRvL..66.2549D,1993tegp.book.....W}. In the case of PSR~J0435+3233, and assuming a near-coplanarity of the inner and outer orbits, this has roughly the same magnitude as in the case of PSR~J0337+1715\footnote{Wex, private communication}. However, the larger projected semi-major axis of the inner orbit ($x = 8.0\, \rm s$ for PSR~J0435+3233 compared to 1.2 s for PSR~J0337+1715) allows for the detection of correspondingly smaller fractional changes of the shape, size and orientation of the inner orbit, which would imply, assuming similar timing precision, that PSR~J0435+3233 could provide an even better test of UFF violation. A detailed evaluation of the potential of the system for tests of gravity theories will be carried out elsewhere, but  its is already clear this is a promising system for such tests.

Note added during the revision: After submitting our paper, we became aware that a different team had reached, independently, very similar conclusions to ours \citep{2026ApJ..1010L...9Y}.

\begin{acknowledgements}
We thank Norbert Wex for many valuable discussions, for providing a preliminary determination of the outer orbit from the equations of \cite{1997ApJ...479..948J} and \cite{2017MNRAS.468.2114P} and for his comments on the manuscript. We also thank Michael Kramer and the anonymous referee for helpful comments on the manuscript. BWS thanks Michael Keith for valuable discussions. We thank David Smith, Pablo Saz Parkinson, Xian Hou, and Matthew Kerr, for reviewing the paper on behalf of the Fermi-LAT collaboration.
The \textit{Fermi} LAT Collaboration acknowledges generous ongoing support
from a number of agencies and institutes that have supported both the
development and the operation of the LAT as well as scientific data analysis.
These include the National Aeronautics and Space Administration and the
Department of Energy in the United States, the Commissariat \`a l'Energie Atomique
and the Centre National de la Recherche Scientifique / Institut National de Physique
Nucl\'eaire et de Physique des Particules in France, the Agenzia Spaziale Italiana
and the Istituto Nazionale di Fisica Nucleare in Italy, the Ministry of Education,
Culture, Sports, Science and Technology (MEXT), High Energy Accelerator Research
Organization (KEK) and Japan Aerospace Exploration Agency (JAXA) in Japan, and
the K.~A.~Wallenberg Foundation, the Swedish Research Council and the
Swedish National Space Board in Sweden.

Additional support for science analysis during the operations phase is gratefully 
acknowledged from the Istituto Nazionale di Astrofisica in Italy and the Centre 
National d'\'Etudes Spatiales in France. This work performed in part under DOE 
Contract DE-AC02-76SF00515.
\end{acknowledgements}

\bibliographystyle{aa}

\bibliography{references.bib} 

\begin{appendix}

\onecolumn
\section{Full timing results}
\label{app:results}
\begin{figure*}[h]
    \centering
    {\includegraphics[width=\textwidth]{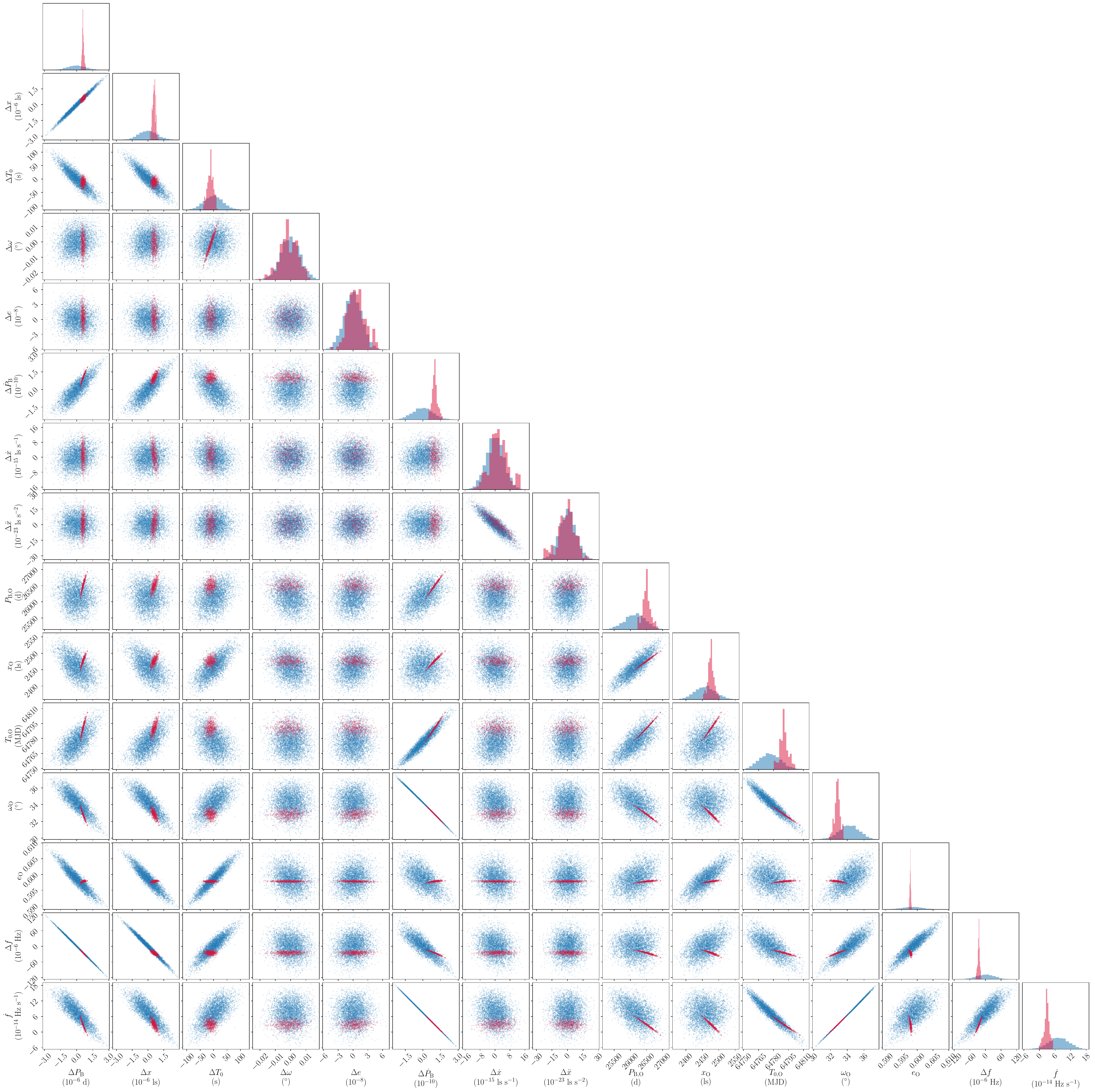}}
     \caption{Corner plot showing all timing parameters, in the same format as Fig.~\ref{fig:corner}. Parameters with labels prefixed by $\Delta$ are offsets relative to the values reported in the left column of Table~\ref{tab:solution}.
     \label{fig:full_corner}}
\end{figure*}

Fig.~\ref{fig:distances} in the main text shows the posterior distribution for the outer-orbital parameters and the pulsar's spin-frequency derivative. Fig.~\ref{fig:full_corner} here shows the full posterior distribution for all parameters included in our timing model. 

\end{appendix}

\end{document}